\pdfoutput=1
\documentclass[10pt,preprint,3p,a4paper,onecolumn]{elsarticle}

\makeatletter
\def\ps@pprintTitle{%
 \let\@oddhead\@empty
 \let\@evenhead\@empty
 \def\@oddfoot{}%
 \let\@evenfoot\@oddfoot}
\makeatother

\usepackage[x11names]{xcolor}
\usepackage[utf8]{inputenc}
\usepackage[english]{babel}
\usepackage{amsmath}
\usepackage{amssymb}
\usepackage[hidelinks]{hyperref}
\usepackage{subcaption}
\usepackage{tikz}
	\usetikzlibrary{patterns,calc,decorations.pathmorphing,decorations.markings,positioning,plotmarks}
\usepackage{placeins}
\usepackage{float}
\usepackage{amsthm}
\usepackage{mathrsfs}
\usepackage{enumitem}
\usepackage{booktabs}

\colorlet{is}{DodgerBlue4}
\colorlet{sb}{Chocolate1!80!white}
\colorlet{ac}{DarkOrchid4!70!black}

\renewcommand{\i}{\mathrm{i}}
\newcommand*\ctrmcell[1]{\omit\hfil$\displaystyle#1$\hfil\ignorespaces}

\begin{document}

\begin{frontmatter}
  \title{Analysis and design of nonlinear resonances via singularity theory}
  \author[ccc]{G.I.~Cirillo\corref{cor1}}
  \ead{gic27@cam.ac.be}
  \author[lme]{G.~Habib}
  \ead{giuseppe.habib@ulg.ac.be}
  \author[lme]{G.~Kerschen}
  \ead{g.kerschen@ulg.ac.be}
  \author[ccc]{R.~Sepulchre}
  \ead{r.sepulchre@eng.cam.ac.uk}
  \cortext[cor1]{Corresponding author}
  \address[ccc]{Control Group, Department of Engineering, University of Cambridge, Cambridge, UK.}
  \address[lme]{Space Structures and Systems Lab (S3L), Department of Aerospace and Mechanical Engineering, University of Li\`ege, Li\`ege, Belgium.}

  \begin{abstract}
    Bifurcation theory and continuation methods are well-established tools for the analysis of nonlinear mechanical systems subject to periodic forcing. We illustrate the added value and the complementary information provided by singularity theory with one distinguished parameter. While tracking bifurcations reveals the qualitative changes in the behaviour, tracking singularities reveals how structural changes are themselves organised in parameter space. The complementarity of that information is demonstrated in the analysis of detached resonance curves in a two-degree-of-freedom system.
  \end{abstract}


\end{frontmatter}

\section{Introduction}
\label{sec:introduction}

Nonlinear resonances are a matter of increasing concern in engineering structures, and bifurcation analysis is by now a standard tool in the study of responses of systems subject to periodic forcing. While this type of approach has proven effective, its main limitation lies in that the frequency response is not considered as a whole. Rather, bifurcations are studied for fixed values of the forcing frequency. On the contrary, singularity theory with a distinguished parameter~\cite{golubitsky1985singularities} (singularity theory in the following), identifies changes in a one-parameter bifurcation diagram, for example in a frequency response. This viewpoint provides a useful complement to the classical use of bifurcation analysis in engineering applications, while employing the same numerical methods~\cite[Ch.7]{govaerts2000numerical}.

We adopt the framework developed in~\cite{golubitsky1979theory}, probably the most successful attempt to use methods of singularity theory (in the broader sense of~\cite{golubitsky2012stable}) in the context of bifurcation problems. The use of this theory is classical in the analysis of the Duffing oscillator~\cite{holmes1976bifurcations}. However, to the best of the authors' knowledge, the engineering literature has not pursued the use of such methods in problems of forced oscillations. This is our motivation to highlight the role of singularity theory in engineering applications concerned with forced oscillations.

A concrete application in this paper is the analysis of isolated branches of solutions in the response of a system. Detached resonance curves have been observed for different systems subject to periodic forcing~\cite{Gatti2010591,Alexander2009445,Starosvetsky2009916,Duan200812}, as well as in other applications involving nonlinearities~\cite{Sarrouy2011727,takacs2008isolated}. Their analysis is not straightforward in the classical framework of bifurcation analysis. In contrast, we argue that singularity theory provides a complete answer to the problem by revealing the organising role of a particular singularity, the asymmetric cusp. We illustrate this result on a two-degree-of-freedom system with cubic springs.

The paper starts with the study of a two-degree-of-freedom system in Section~\ref{sec:nltva}, which illustrates and motivates the proposed analysis. Section~\ref{sec:singularity-theory} presents those results of singularity theory that have a direct impact on the analysis of nonlinear frequency responses. Section~\ref{sec:cusp-sing-organ} introduces the role of the asymmetric cusp in organising DRCs in parameter space. Section~\ref{sec:design-5th-spring} builds on the previous analysis to investigate the inclusion of a fifth-order spring to avoid DRCs. Conclusions are presented in Section~\ref{sec:conclusions}.

\section{A motivating example}\label{sec:nltva}

\begin{figure}[!h]
  \centering
  \includegraphics{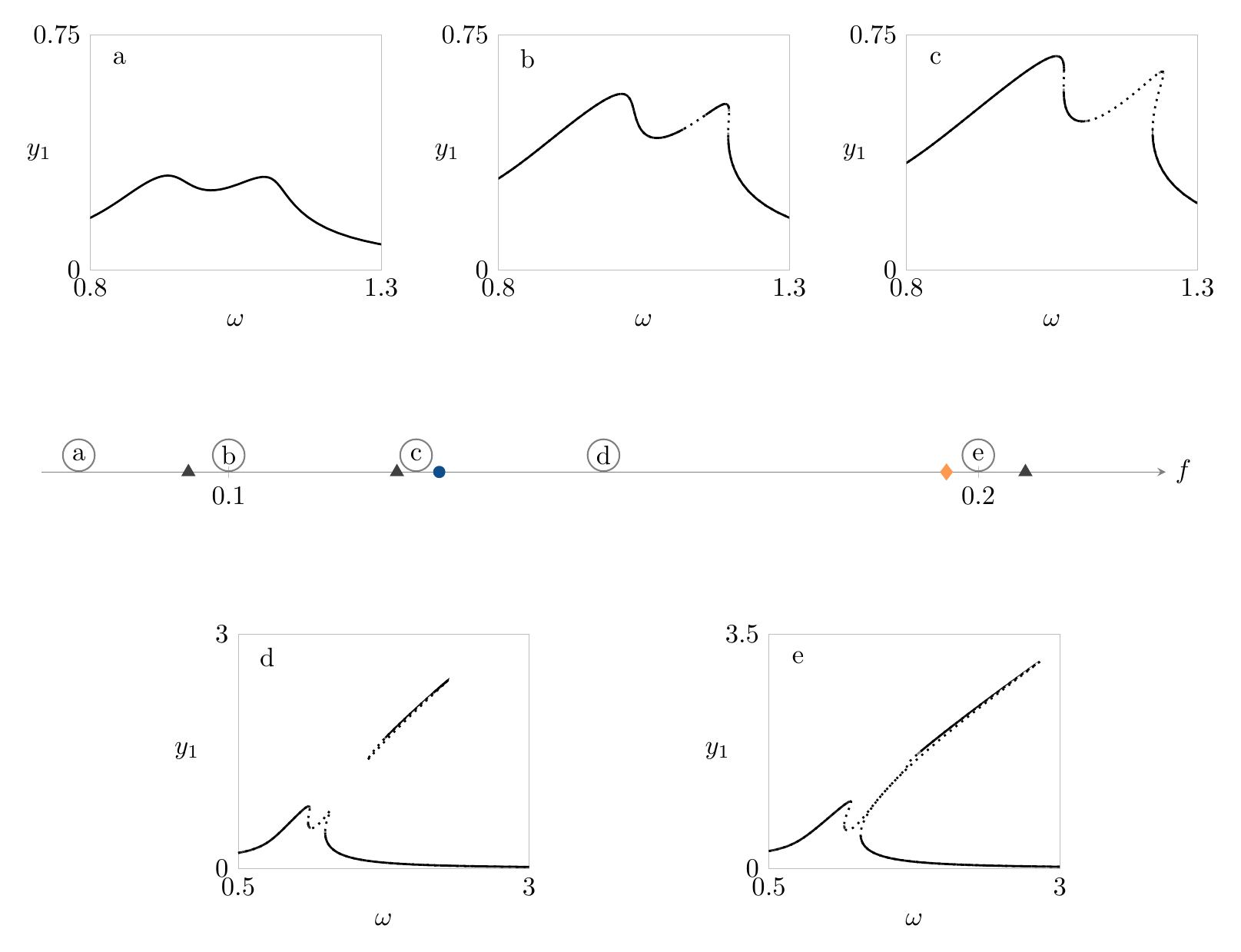}
  \caption[Frequency responses of an NLTVA attached to a Duffing oscillator]{Frequency responses of an NLTVA attached to a Duffing oscillator. Transition between different diagrams correspond to different singularities (%
\tikz \node[darkgray, scale=1.25,inner sep=.75mm] at (0,0) {\nullfont\pgfuseplotmark{triangle*}};~hysteresis, %
\tikz \node[is, scale=1.1,inner sep=.75mm] at (0,0) {\nullfont\pgfuseplotmark{*}};~isola, %
\tikz \node[sb, scale=1.35,inner sep=.6mm] at (0,0) {\nullfont\pgfuseplotmark{diamond*}};~simple bifurcation). Excerpts illustrate the different type of frequency response, the effect of the last hysteresis is not shown as it is minimal. a) $f=0.05$, b) $f=0.1$, c) $f=0.125$, d) $f=0.15$, e) $f=0.2$.}\label{fig:fre-res}
\end{figure}

To motivate the developments proposed in this paper, a harmonically-forced two-degree-of-freedom system possessing two cubic nonlinearities is considered herein:
\begin{equation}\label{eq:NLTVAeom}
    \begin{gathered}
      M \ddot{y}+C\dot{y}+Ky+f_{nl}(y)=g f \cos(\omega t),\\[2ex]
      M=\left( \begin{array}{cc}m_1 & 0 \\ 0 & m_2\end{array}\right),\,\,\,\,\,\,K=\left(\begin{array}{cc}k_1+k_2 & -k_2\\ -k_2 & k_2\end{array}\right),\,\,\,\,\,\,C=\left(\begin{array}{cc}c_1+c_2 & -c_2 \\ -c_2 & c_2\end{array}\right),\\[2ex]
 f_{nl}(y)=\left(\begin{array}{c}\alpha_1y_1^3+\alpha_2(y_1-y_2)^3\\\alpha_2(y_2-y_1)^3\end{array}\right),\,\,\,\,\,\,g=\left(\begin{array}{c}1\\0\end{array}\right),
    \end{gathered}
\end{equation} 
where the parameters are listed in Table~\ref{tab:par}. This system represents a nonlinear primary system to which a nonlinear absorber, termed the nonlinear tuned vibration absorber (NLTVA), is attached. Similarly to what is achieved with the classical linear vibration absorber, the NLTVA can maintain two equal peaks in the frequency response of the primary system, and this despite the frequency-amplitude dependence of nonlinear oscillations (\cite{Habib2015nonlinear}, see also the insets in Fig.~\ref{fig:fre-res}). However, one important challenge is that nonlinear systems can exhibit complex and rich dynamics. For instance, Fig.~\ref{fig:fre-res}d  depicts a detached resonance curve (DRC) that is clearly detrimental to the performance of the nonlinear device.

\begin{figure}[!ht]
  \centering
  \includegraphics{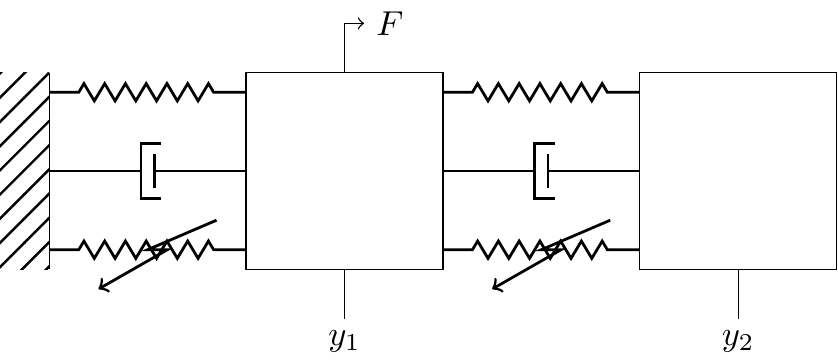}
  \caption{Nonlinear tuned vibration absorber attached to a primary system.}\label{fig:SysScheme}
\end{figure}

\begin{table}[h]
\centering
\begin{tabular}{l l l}
\toprule
 & primary system  & absorber \\
\midrule
mass & $m_1=1$ & $m_2=.05$ \\
linear stiffness & $k_1=1$ & $k_2=0.0454$ \\
linear damping & $c_1=0.002$ & $c_2=0.0128$ \\
nonlinear stiffness & $\alpha_1=1$ & $\alpha_2=0.0042$ \\
\bottomrule
\end{tabular}
\caption{Parameters primary system and absorber}\label{tab:par}
\end{table}

This adverse dynamics was thoroughly investigated using the numerical continuation of periodic solutions and their bifurcations in~\cite{Detroux2015performance}. Here we complement this analysis with the viewpoint of singularity theory. In the same spirit of what was proposed in~\cite{zhang2011periodically,wiser2015bifurcation} for forced Hopf bifurcations, we apply singularity theory to a bifurcation diagram identified to the frequency response of the system. The starting point is to reduce the frequency response of~\eqref{eq:NLTVAeom} to a scalar equation; to obtain it we use the harmonic balance method retaining one harmonic. The results obtained with this approximation are then verified through numerical computation of frequency responses using an algorithm similar to the one used in~\cite{Peeters2009nonlinear} for the computation of nonlinear normal modes.~\ref{sec:HBNLTVA} details the reduction of the harmonic balance equations to a scalar equation
\begin{equation}
  \label{eq:aeq}
  g(x,\omega,f,k_2,\alpha_2,c_2)=0,
\end{equation}
where $x=y_1-y_2$ is the relative displacement, called the state variable in the terminology of~\cite{golubitsky1985singularities}. The frequency $\omega$ is the bifurcation (distinguished) parameter. 

The outcome of applying singularity theory to~\eqref{eq:aeq} is summarised in Fig~\ref{fig:fre-res}. At low forcing amplitudes the response resembles that of a linear system (Fig~\ref{fig:fre-res}a), but as $f$ is increased parts of it become bistable (Fig.~\ref{fig:fre-res}b and c, note that part of these responses is unstable due to the appearance of quasiperiodic motions, see~\cite{Detroux2015performance} for further details). These bistable regions are delimited by two fold points, which appear in correspondence of a hysteresis singularity. Other types of singularities are encountered increasing further the forcing amplitude: An isola corresponds to the appearance of a DRC (Fig.~\ref{fig:fre-res}d), while a simple bifurcation marks the merging of this isolated branch of solutions with the main one (Fig.~\ref{fig:fre-res}e). Thus singularities define the transitions between the qualitatively different frequency responses shown in the figure, thereby providing a global perspective on the dynamics that may be encountered when the forcing amplitude is increased.


Clearly, the two theoretical frameworks, bifurcation analysis and singularity theory, provide distinct information. Continuation methods of bifurcation theory primarily track fold bifurcations (or limit points in the terminology of~\cite{golubitsky1985singularities}), characterised by the two algebraic conditions
\begin{equation}
  \label{eq:lim-fold-cond}
  g=0, \quad \frac{\partial g}{\partial x}=0.
\end{equation}
These two conditions are the defining conditions of a singularity, but the transition points identified on the axis in Fig.~\ref{fig:fre-res} are characterised by one of the following three further degeneracies:
\begin{itemize}
\item hysteresis (\tikz \node[darkgray, scale=1.25,inner sep=.75mm] at (0,0) {\nullfont\pgfuseplotmark{triangle*}};)
\begin{equation}\label{eq:hyst-def}
  \frac{\partial^2 g}{\partial x^2}=0, \quad \frac{\partial g}{\partial \omega} \neq 0, \quad \frac{\partial^3 g}{\partial x^3} \neq 0;
\end{equation}
\item isola (\tikz \node[is, scale=1.1,inner sep=.75mm] at (0,0) {\nullfont\pgfuseplotmark{*}};)
\begin{equation}\label{eq:isola-def}
  \frac{\partial g}{\partial \omega}=0, \quad \frac{\partial^2 g}{\partial x^2} \neq 0, \quad \det(d^2g) > 0;
\end{equation}
\item simple bifurcation (\tikz \node[sb, scale=1.35,inner sep=.6mm] at (0,0) {\nullfont\pgfuseplotmark{diamond*}};)
\begin{equation}\label{eq:sim-nondeg}
  \frac{\partial g}{\partial \omega}=0, \quad \frac{\partial^2 g}{\partial x^2} \neq 0, \quad \det(d^2g) < 0;
\end{equation}
\end{itemize}
where $d^2g$ is the Hessian matrix of $g(x,\omega)$.

All the transitions in the frequency response of the NLTVA~\eqref{eq:NLTVAeom} are defined by one of these three singularities, a result that follows from the basic theory recalled in the next section.

\section{Singularities of codimension one and transition set}
\label{sec:singularity-theory}

Singularity theory with one distinguished parameter is the study of bifurcation diagrams defined by the equation
\begin{equation}
  \label{eq:base_eq}
  g(x,\omega,\mu)=0,
\end{equation}
where $x$ is the state variable, $\omega$ is the bifurcation parameter (the frequency in the present work) and $\mu\in\mathbb{R}^n$ are additional parameters. We review the basic elements of the theory necessary to understand its application to the NLTVA. For a detailed exposition the reader is referred to~\cite{golubitsky1985singularities,golubitsky1988singularities}.

\subsection{Singularities of codimension one}
\label{sec:sing-codim-one}

This section starts by analysing more in details the three singularities encountered in Section~\ref{sec:nltva}: hysteresis, isola and simple bifurcation. As any other singularities, these are characterised by two groups of conditions on the derivatives of~\eqref{eq:base_eq} at a point. \emph{Defining conditions}, which corresponds to derivatives of~\eqref{eq:base_eq} being equal to zero, and \emph{Nondegeneracy conditions} corresponding (in the simpler cases) to derivatives of~\eqref{eq:base_eq} being different from zero. For example the defining conditions for the hysteresis are
\begin{equation}
  \label{eq:hyst-def-cond}
  g=\frac{\partial g}{\partial x}=\frac{\partial^2 g}{\partial x^2}=0,
\end{equation}
while the nondegeneracy conditions are
\begin{equation}
  \label{eq:hyst-nondeg-cond}
  \frac{\partial g}{\partial \omega}\neq 0, \qquad \frac{\partial^3 g}{\partial x^3}\neq 0.
\end{equation}

The defining conditions always include those in~\eqref{eq:lim-fold-cond}, corresponding to the failure of the implicit function theorem. However, these alone do not imply a qualitative change in a bifurcation diagram. In fact, generically~\eqref{eq:lim-fold-cond} indicates the presence of a fold, which is a \emph{persistent} singularity, i.e. it is preserved by small perturbations. To obtain a qualitative change in a diagram a \emph{nonpersistent} singularity is necessary, one that due to small perturbations (for example due to a small change of a parameter value) disappears, leading to different possibilities.

\begin{figure}[!h!]
  \centering
  \catcode`$=3 
  \begin{subfigure}[c]{.3\linewidth}
    \centering
    \includegraphics{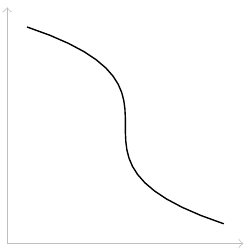}
    \caption{Unperturbed diagram.}\label{fig:hyst-un}
  \end{subfigure}%
  \begin{subfigure}[c]{.6\linewidth}
    \centering
    \includegraphics{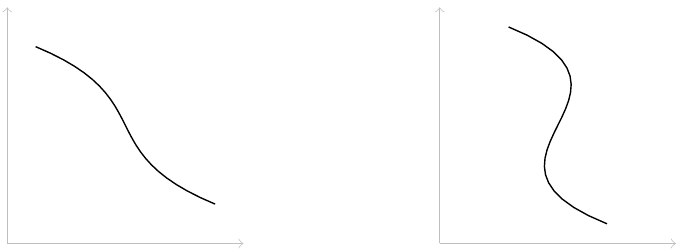}
    \caption{Perturbed diagrams.}\label{fig:hyst-per}
  \end{subfigure}%
  \caption{The hysteresis and its perturbations.}
  \label{fig:hyst}
\end{figure}

The first example of nonpersistent singularity is the hysteresis. Geometrically these points are characterised by vertical tangent, as in Fig.~\ref{fig:hyst-un}. When perturbed, the nonpersistent diagram can only result in one of the two diagrams in Fig.~\ref{fig:hyst-per}.

\FloatBarrier

\begin{figure}[!h]
  \centering
  \begin{subfigure}[c]{.3\linewidth}
    \centering
    \includegraphics{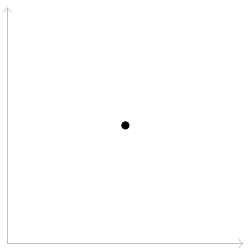}
    \caption{Unperturbed diagram.}\label{fig:isola-un}
  \end{subfigure}%
  \begin{subfigure}[c]{.6\linewidth}
    \centering
    \includegraphics{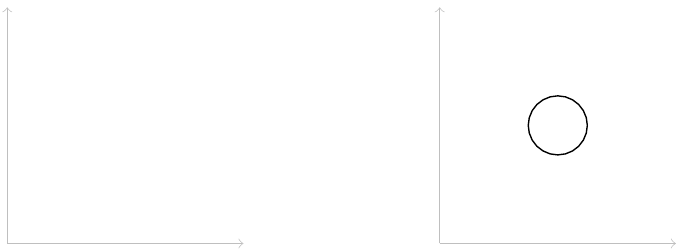}
    \caption{Perturbed diagrams.}\label{fig:isola-per}
  \end{subfigure}%
  \caption{The isola and its perturbations.}
  \label{fig:isola}
\end{figure}

The second example of nonpersistent singularity is the isola, which was found at the transition from Fig.~\ref{fig:fre-res}c to~\ref{fig:fre-res}d, when a DRC appears. This corresponds to the presence of an isolated solution as in Fig.~\ref{fig:isola-un}. When perturbed, this nonpersistent diagram can only result in two outcomes: either no solutions or a closed branch of solutions, both shown in Fig.~\ref{fig:isola-per}.

\begin{figure}[!h]
  \centering
  \begin{subfigure}[c]{.3\linewidth}
    \centering
    \includegraphics{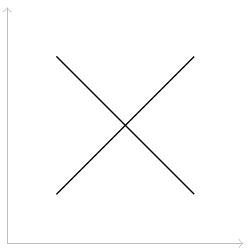}
  \caption{Unperturbed diagram.}\label{fig:sim-un}
  \end{subfigure}%
  \begin{subfigure}[c]{.6\linewidth}
    \centering
    \includegraphics{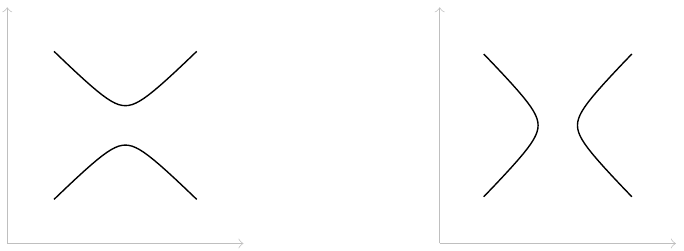}
    \caption{Perturbed diagrams.}\label{fig:sim-per}
  \end{subfigure}%
  \caption{The simple bifurcation and its perturbations.}
  \label{fig:sim-bif}
\end{figure}

The third and last example of nonpersistent singularity is the simple bifurcation point, which corresponds to the centre of X-shaped diagrams, as illustrated in Fig.~\ref{fig:sim-un}. The corresponding perturbations are shown in Fig.~\ref{fig:sim-per}. Clearly, this is the local phenomenon that underlies the merging of a DRC with a main branch.

The three singularities above account for all the codimension one singularities. The codimension of a singularity can be thought of as a measure of its complexity, and it is given by the number of defining conditions that characterise it beyond the two conditions~\eqref{eq:lim-fold-cond}. Using this concept, singularities can be classified. All singularities up to codimension three are presented in~\cite[Ch.4]{golubitsky1985singularities}, while in~\cite{keyfitz1986classification} the classification is extended up to codimension seven.

The codimension coincides also with the number of parameters necessary to obtain all the possible perturbations of a given singularity, i.e. the number of parameters in a universal unfolding. Without entering into details, for which the reader is referred to~\cite[Ch.3]{golubitsky1985singularities}, a universal unfolding is the technical tool used to enumerate the persistent diagrams that can be obtained by small perturbations of a singular one, as in the case of Fig.~\ref{fig:hyst},~\ref{fig:isola} and~\ref{fig:sim-bif}.

\subsection{Persistent and nonpersistent diagrams}
\label{sec:pers-nonp-diagr}

The previous section covered all the codimension one singularities. Because they are nonpersistent phenomena, they are difficult to observe in numerical simulations or during experiments. Their importance lies in that they act as boundaries between different persistent diagrams, those observed in simulations and experiments. It is thus clear that the concept of persistence has a central role in obtaining the different possible solutions of~\eqref{eq:base_eq}. To give a precise definition of nonpersistence it is necessary to introduce the concept of equivalence (also referred as contact equivalence). Two diagrams $g(x,\omega)=0$ and $f(x,\omega)=0$ are said to be equivalent if there exists a diffeomorphism $\Phi(x,\omega)=(\phi_x(x,\omega),\phi_\omega(\omega))$ and a function $S(x,\omega)$ such that
\begin{equation}
  \label{eq:contactEquivalence}
  f(x,\omega)=S(x,\omega)g(\phi_x(x,\omega),\phi_\omega(\omega)),
\end{equation}
with $S(x,\omega)>0$, $\frac{\partial \phi_x}{\partial x}>0$ and $\frac{d\phi_\omega}{d\omega}>0$ (these last conditions guarantee that stability properties are preserved under equivalence, for geometric aspects as those considered in this work these conditions could be relaxed~\cite{govaerts2000numerical}). To complete the definition other conditions are imposed in~\cite{golubitsky1985singularities}. These depend from the context in which~\eqref{eq:contactEquivalence} is used and lead to two distinct (although related) concepts: global and local equivalence. As the names suggest, the first one considers functions which are globally defined, for which $x$ and $\omega$ are supposed to vary in a fixed set $U\times L$. In this case~\eqref{eq:contactEquivalence} has to additionally preserve this domain. In the case of local equivalence the functions involved are supposed to be defined only in a neighbourhood of the origin, more precisely, instead of functions \emph{germs} are considered. These are defined as equivalence classes of functions, and additional conditions assure that equivalence~\eqref{eq:contactEquivalence} is well-defined for these classes. The precise definitions are not necessary here, as more technical aspects of the theory for which this difference is more relevant are not considered. However, it is necessary to distinguish which type of equivalence is used to obtain a result. For example classifications of perturbations, as those in Fig.~\ref{fig:hyst-per},~\ref{fig:isola-per} and~\ref{fig:sim-per}, are all local results, while the theorem stated in the following uses the global version of equivalence.

Having defined equivalence, it is possible to define persistence: a diagram is said to be persistent if sufficiently small perturbations result in equivalent diagrams, and it is nonpersistent otherwise. Referring to the previous section, all the frequency responses shown are persistent. Moreover, all the responses that correspond to values of $f$ between two consecutive singularities in Fig.~\ref{fig:fre-res} are equivalent. This gives a more precise idea of how singularity theory can be used to analyse forced oscillations: it provides the tools to divide the parameter space in different zones characterised by a unique (up to equivalence) response. A noteworthy result in this direction is~\cite[Ch.3~ Thm.10.1]{golubitsky1985singularities}, which gives the possible sources of nonpersistence for a parametrised family of diagrams $g(x,\omega,\mu)$ where $\mu\in W$, with $W\subset\mathbb{R}^k$ a disk. Specifically, there are three sources of nonpersistence (that can occur in the interior of the domain considered), corresponding to three sets:
\begin{enumerate}[label=\roman*]
\item simple bifurcation and isola points,
\[\mathscr{B}=\left\{\mu\in\mathbb{R}^k:\exists(x,\omega)\in\mathbb{R}\times\mathbb{R} \text{ such that } g=\frac{\partial g}{\partial x}=\frac{\partial g}{\partial \omega}=0 \text{ at } (x,\omega,\mu) \right\};\]
\item hysteresis points,
\[\mathscr{H}=\left\{\mu\in\mathbb{R}^k:\exists(x,\omega)\in\mathbb{R}\times\mathbb{R} \text{ such that } g=\frac{\partial g}{\partial x}=\frac{\partial^2 g}{\partial x^2}=0 \text{ at } (x,\omega,\mu) \right\};\]
\item double limit points,
\[\mathscr{D}=\left\{\mu\in\mathbb{R}^k:\exists(x_1,x_2,\omega)\in\mathbb{R}\times\mathbb{R}\times\mathbb{R}, x_1\neq x_2, \text{ such that } g=\frac{\partial g}{\partial x}=0 \text{ at } (x_i,\omega,\mu),\, i=1,2 \right\}.\]
\end{enumerate}
Based on these the \emph{transition set} is defined as $\Sigma=\mathscr{B}\cup\mathscr{H}\cup\mathscr{D}$. The theorem states that diagrams obtained for $\mu$ in the same connected component of $W\setminus\Sigma$ are equivalent. 

\begin{figure}[h]
  \centering
  \begin{subfigure}[c]{.3\linewidth}
    \centering
    \includegraphics{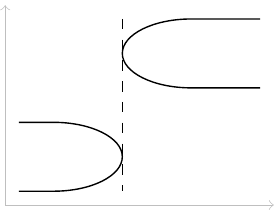}
    \caption{Unperturbed diagram.}\label{fig:dlp-un}
  \end{subfigure}%
  \begin{subfigure}[c]{.6\linewidth}
    \centering
    \includegraphics{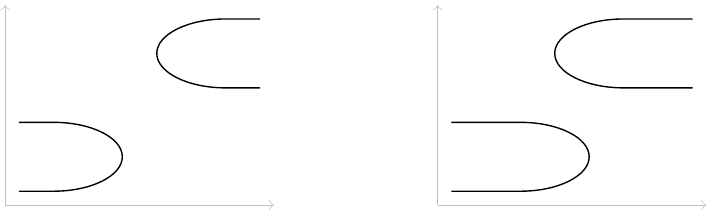}
    \caption{Perturbed diagrams.}\label{fig:dlp-per}
  \end{subfigure}%
  \caption{The double limit point and its perturbations.}
  \label{fig:double-lim-pts}
\end{figure}

The conditions defining $\mathscr{H}$ correspond to the defining conditions of the hysteresis, while those defining $\mathscr{B}$ correspond to isola and simple bifurcations. Therefore, when crossing one of these two sets, we expect one of the phenomena described in the previous section to take place. This is because the nondegeneracy conditions, which correspond to inequalities, are more likely to be satisfied than not. Of course, for specific values of the parameters this might not be true, in which case a more degenerate singularity is present. The set $\mathscr{D}$ is partially different from the other two, in that it corresponds to a nonlocal phenomenon, the occurrence of two limit points for the same value of $\omega$ (see Fig.~\ref{fig:double-lim-pts}).

For completeness we mention that other sources of nonpersistence are possible since $x$ and $\omega$ vary in a bounded domain $U\times L$. These are related to how the diagram meets the boundary of the domain and are discussed in~\cite[Ch.3~\S 10]{golubitsky1985singularities}. Since the boundary of frequency and amplitude are not fixed in most applications, but can be moved to avoid these types of phenomena, these are unlikely to be important in the present case and thus not considered.

\subsection{Numerical computation}
\label{sec:numer-comp}

The knowledge of all the possible sources of nonpersistence can be used as a base for algorithms that explore the parameter space of a system, and methods based on it appeared in the context of chemical reactions in~\cite{balakotaiah1981analysis,balakotaiah1984global}. Ideally, one would compute the transition set $\Sigma$ and then obtain a rather complete picture of the possible frequency responses. However, even if from a theoretical viewpoint the three components of $\Sigma$ are easily defined, their construction in applications might not be straightforward. One major drawback is that the conditions defining the three sets are given in terms of a scalar function $g(x,\omega)$. While techniques exist to reduce the problem of computing a frequency response to the solution of a set of nonlinear equations in $x$ and $\omega$ (for example through harmonic balance~\cite{Detroux2015harmonic}), reducing the whole system to only one equation is not always possible. 

\begin{figure}[!h]
  \centering
  \includegraphics{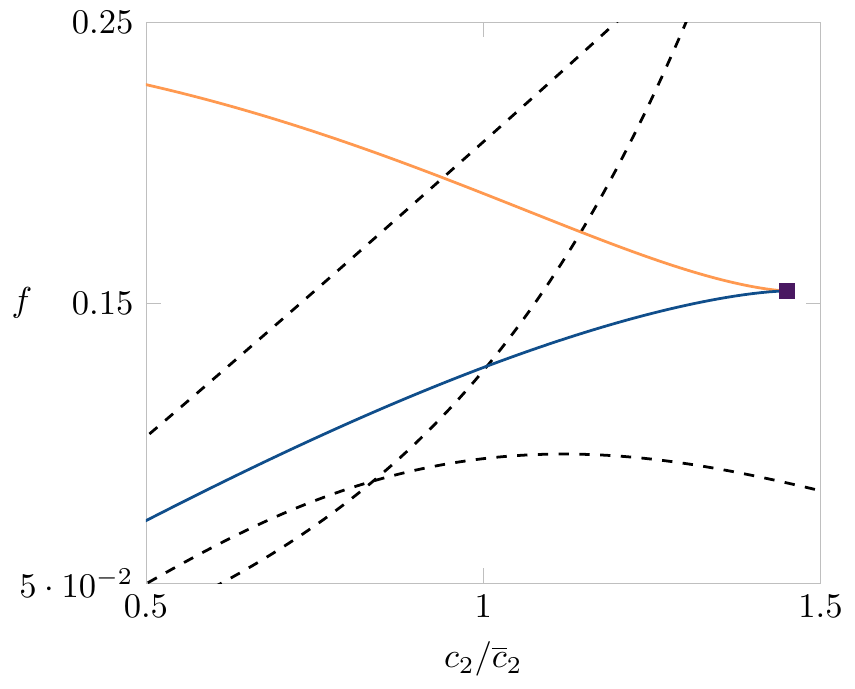}
  \caption[Transition set for an NLTVA attached to a Duffing oscillator]{Transition set for an NLTVA attached to a Duffing oscillator. The parameter space is divided into different zones, each characterised by one frequency response (up to equivalence) (\tikz \draw[dashed,thick] (0,0)--(15pt,0) node {};~hysteresis $(\mathscr{H})$, \tikz \draw[is,thick] (0,0)--(15pt,0) node {};~isola $(\mathscr{B}')$, \tikz \draw[sb,thick] (0,0)--(15pt,0) node {};~simple bifurcation $(\mathscr{B})$).}
  \label{fig:tranVar}
\end{figure}

It turns out that this reduction might not be necessary in view of developing a general algorithm, as equivalent conditions on a full system can be derived. This is a consequence of the fact that the conditions defining the sets $\mathscr{B}$ and $\mathscr{H}$ correspond to the presence of a singularity, and those defining $\mathscr{D}$ correspond to the occurrence of two fold points for the same value of $\omega$. All these phenomena can be identified on a full system of equations (see~\cite{govaerts2000numerical}). Following this idea, the transition set could be constructed tracking singularities, in a similar fashion to what is possible for bifurcations~\cite{Detroux2015harmonic,dhooge2003matcont}. 

In the following this idea is used in its simpler form. Because we only consider situations in which the frequency response can be approximated with one equation, we only need to track singularities, which is achieved by solving the defining conditions of a singularity for different values of a parameter (for the numerical solution of equations the python package SciPy~\cite{scipy} was used).

It is worth mentioning that the exposition above only considers a scalar unknown $x$ whereas methods of singularity theory apply to $n$-dimensional problems $x\in\mathbb{R}^n$~\cite[Ch.9]{golubitsky1985singularities}; this generalisation, however, does not add much to the present discussion, and it is therefore not included.

Fig.~\ref{fig:tranVar} shows the result of the procedure outlined above for system~\eqref{eq:NLTVAeom}. Starting from the points in Fig.~\ref{fig:fre-res} and varying $c_2$, the parameter space is divided into regions characterised by a unique frequency response with respect to equivalence~\eqref{eq:contactEquivalence}. The set $\mathscr{B}$ can be divided into two parts, corresponding to isola singularities and simple bifurcations, which delimit the values of $f$ for which a DRC is present. Similarly, crossing one of the three lines composing the set $\mathscr{H}$ results in the merging (or generation) of two fold points.

We mention that some more quantitative aspects of the response are not captured by this analysis. For example, frequency responses corresponding to higher values of $c_2$ contain a unique resonance peak, but are equivalent to those found for lower values of $c_2$ and $f$, which contain two peaks. While the fact that the amplitude corresponding to two hysteresis points grows and exceeds the limits considered in figure suggests a strong change, the disappearance of one resonance peak cannot be inferred directly. Hence, the proposed analysis is complementary to bifurcation analysis, singularity theory provides a qualitative picture of the different possibilities, which can be later refined.

\section{A cusp singularity organises detached resonance curve}
\label{sec:cusp-sing-organ}

A central insight in singularity theory is that more degenerate singularities organise transitions between parameter regions with a lower order of degeneracy. Hence, distinct regions containing fold points in Fig.~\ref{fig:fre-res} are separated by few codimension one singularities. This approach can be repeated with singularities of higher codimension.

We illustrate its value in analysing the detached resonance curves of NLTVA. Figure~\ref{fig:fre-res} indicates that this regime of the frequency response is delimited by an isola (\tikz \node[is, scale=1.1,inner sep=.75mm] at (0,0) {\nullfont\pgfuseplotmark{*}};) and a simple bifurcation (\tikz \node[sb, scale=1.35,inner sep=.6mm] at (0,0) {\nullfont\pgfuseplotmark{diamond*}};). In turn, Fig.~\ref{fig:tranVar} reveals that the sets $\mathscr{B}$ and $\mathscr{B}'$ merge for $c_2^*=1.449\overline{c}_2$ ($\overline{c}_2$ nominal value) and that no DRC exists for $c_2>c_2^*$. The value of $c_2^*$ is in good agreement with the one found in~\cite{Detroux2015performance} ($\approx 1.44\overline{c}_2$) considering that here only one harmonic is retained. Singularity theory identifies this critical point as a singularity of codimension two, the asymmetric cusp.

\subsection{The asymmetric cusp}
\label{sec:asym-cusp}

The defining conditions of the asymmetric cusp are
\begin{equation}
  g=\frac{\partial g}{\partial x}=\frac{\partial g}{\partial \omega}=\det(d^2g)=0,
\end{equation}
while the nondegeneracy conditions are
\begin{equation}
  \frac{\partial^2 g}{\partial x^2} \neq 0, \qquad \frac{\partial^3 g}{\partial v^3} \neq 0,
\end{equation}
where $\frac{\partial}{\partial v}$ is the directional derivative with respect to a zero eigenvector $v$ of $d^2g$, the Hessian of $g(x,\omega)$.

\begin{figure}[!h]
  \centering
  \includegraphics{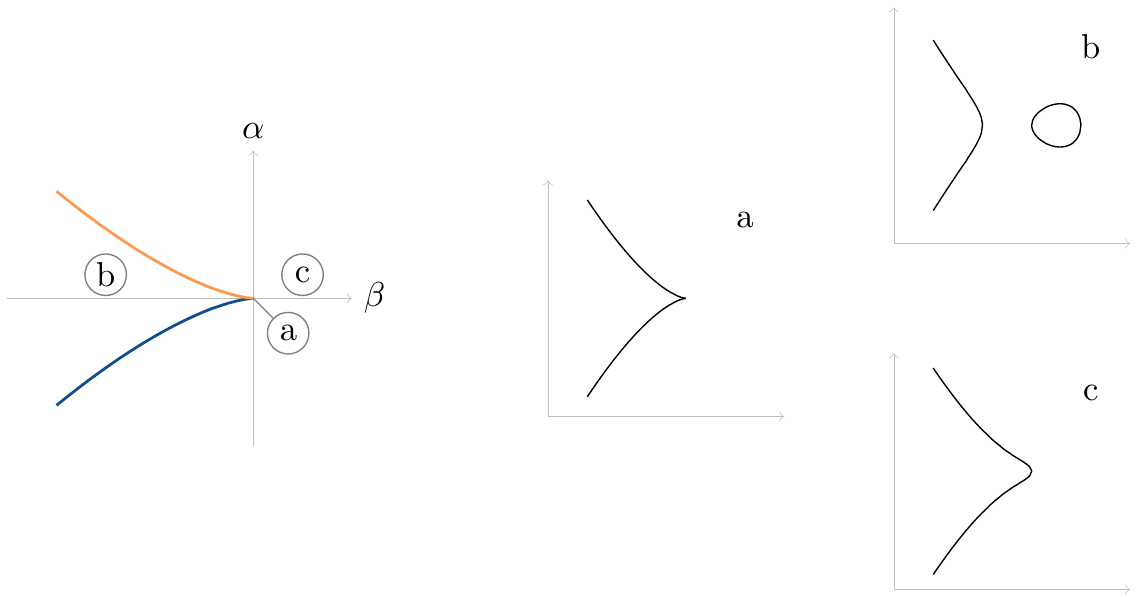}
  \caption{The asymmetric cusp and its persistent perturbations. a) unperturbed diagram; b-c) persistent perturbations.}
  \label{fig:asym-pert}
\end{figure}

Fig.~\ref{fig:asym-pert}a shows the (nonpersistent) diagram corresponding to the normal form of this singularity
\begin{equation}
  \label{eq:nf_as}
  g(x,\omega)=x^2+\omega^3=0,
\end{equation}
while Fig.~\ref{fig:asym-pert}b and c illustrate the persistent diagrams of the universal unfolding
\begin{equation}
  \label{eq:un_as}
  G(x,\omega,\alpha,\beta)=x^2+\omega^3+\alpha+\beta\omega=0.
\end{equation}

The bifurcation set in Fig.~\ref{fig:asym-pert} is easily recognised in the NLTVA bifurcation set in Fig.~\ref{fig:tranVar}. Indeed, whenever an asymmetric cusp is found, singularity theory predicts a line of simple bifurcation points and a line of isola points, delimiting the presence of a separate branch of solutions. In the context of forced oscillations this clearly delimits the presence of DRCs. In other words the asymmetric cusp organises the presence of DRCs in parameter space.

\subsection{Delimiting the appearance of a DRC}
\label{sec:delim-appe-drc}

\begin{figure}[!t]
  \centering
  \includegraphics{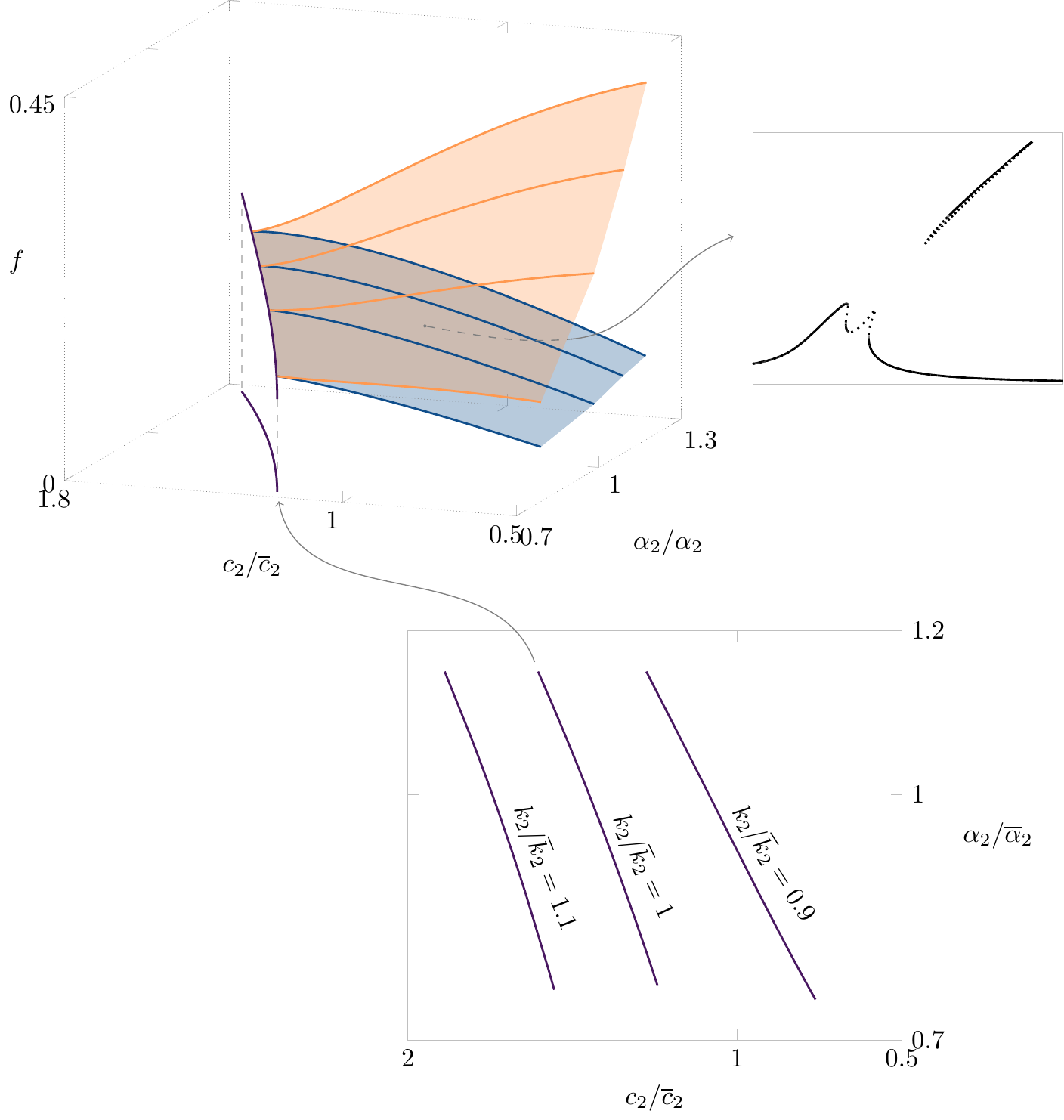}
  \caption[Characters of DRCs in parameter space]{Characters of DRCs in parameter space: bottom plot, lines of asymmetric cusps in the $c_2$-$\alpha_2$ plane for different values of $k_2$; each line determines the boundary in parameter space of the region where a DRC is present, as shown in the upper plots (\tikz \draw[ac,thick] (0,0)--(15pt,0) node {};~asymmetric cusp, \tikz \draw[is,thick] (0,0)--(15pt,0) node {};~isola, \tikz \draw[sb,thick] (0,0)--(15pt,0) node {};~simple bifurcation).}
  \label{fig:drc-par}
\end{figure}

The starting point of our analysis is Fig.~\ref{fig:tranVar}, and specifically the asymmetric cusp (\tikz \node[scale=1.2,ac,inner sep=.75mm] at (0,0) {\pgfuseplotmark{square*}};). The points corresponding to this singularity describe a line in the $\alpha_2$-$c_2$-$f$ space, from which it is possible to construct the entire set $\mathscr{B}$ and delimit the presence in parameter space of a DRC. Looking at Fig.~\ref{fig:tranVar} and considering that increasing $c_2$ is likely to destroy  this phenomenon, it is not difficult to understand how this line of asymmetric cusps can be used to delimit the presence of a DRC. Starting from a point on this line, if $c_2$ is decreased a detached curve is found, while no DRC is present for higher values of the damping, as shown in the 3D plots of Fig.~\ref{fig:drc-par}. This picture agrees, at least at the qualitative level, with the analysis of~\cite{Detroux2015performance}, where it was found for $c_2$ fixed at its nominal value, that as $\alpha_2$ increases so do the values of forcing at which the DRC appears and merges with the main branch.

The analysis is further developed in Fig.~\ref{fig:drc-par}. Projecting the line of asymmetric cusps onto the $\alpha_2$-$c_2$ space the zone in which a DRC is present can be delimited for different values of $k_2$, forming a surface in the $\alpha_2$-$c_2$-$k_2$ space. The plot in the bottom shows for few values of $k_2$ the lines of asymmetric cusp points, which delimit the presence of a DRC, as illustrated in the 3D plot. Hence, through this figure it is possible to determine the values of $\alpha_2$, $k_2$ and $c_2$ (that is, all design parameters) that allow for a DRC for some forcing amplitude. In that sense, singularity theory provides a full characterisation of DRCs in the NLTVA.

\section{Design of a fifth-order spring}
\label{sec:design-5th-spring}

Detached resonance curves might severely restrict the application range of a nonlinear vibration absorber. To further illustrate the potential value of singularity theory in engineering applications, we show in this section how DRCs can be eliminated by means of a fifth order spring.

Specifically, we revisit the analysis in the previous sections with the modified characteristic
\begin{equation}
  \label{eq:fif_nlf}
  f_{nl}(y) = \left( \begin{array}{c} \alpha_1y_1^3+\alpha_2(y_1-y_2)^3+\beta(y_1-y_2)^5 \\ \alpha_2(y_2-y_1)^3+\beta(y_2-y_1)^5 \end{array} \right).
\end{equation}

\begin{figure}
  \centering
  \includegraphics{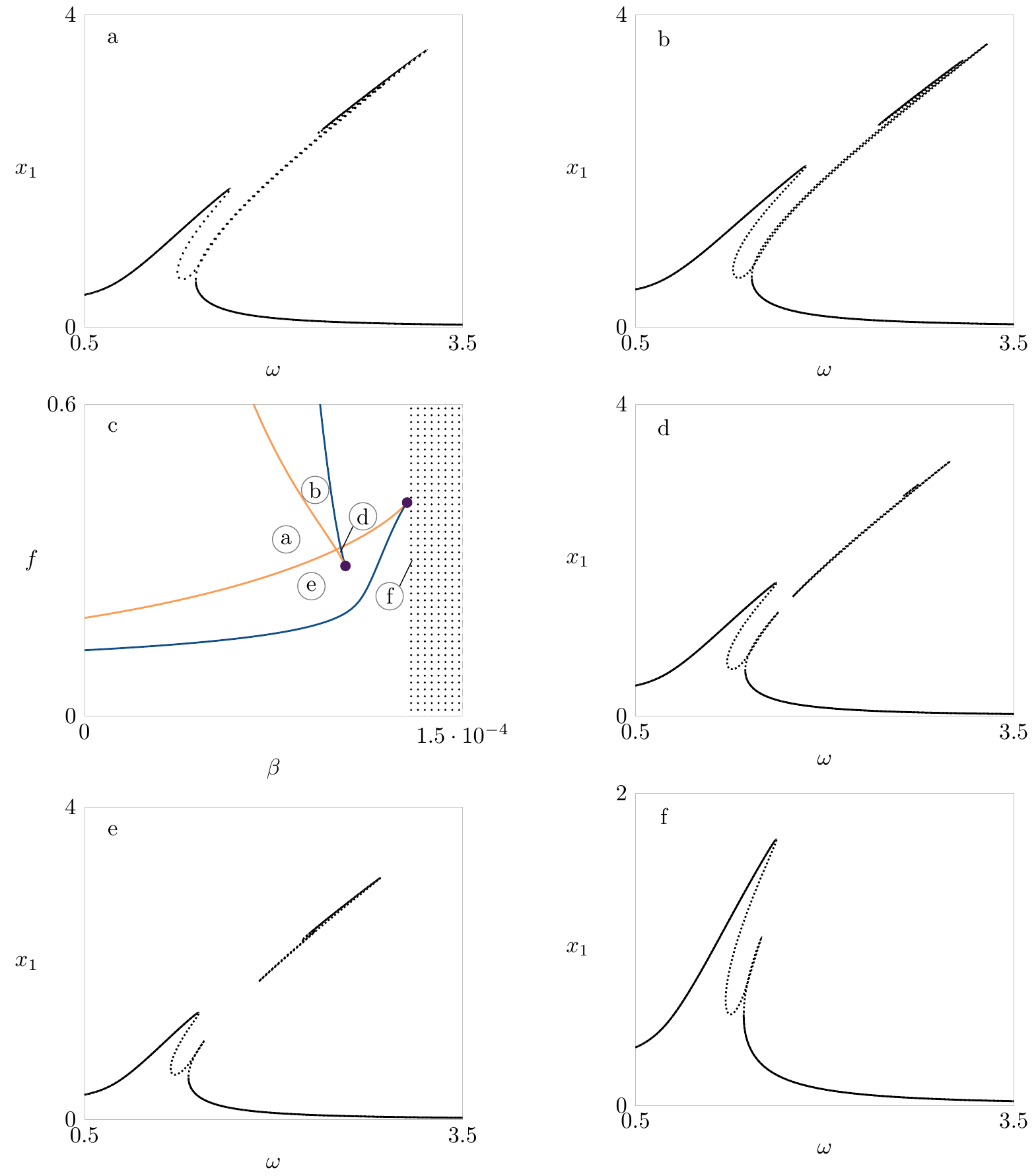}
  \caption{When a fifth order spring is present appearance and merging of DRCs are organised in the parameter space by two asymmetric cusps (purple dot in c), this allows one to identify the dotted area, in which no DRC is present for any real value of $f$; case b and d correspond to the appearance of a new DRC, in Fig.~\ref{fig:x2fr} the corresponding responses for the second mass are reported; \quad a) $\beta=8\,10^{-5}$, $f=.34$;\quad b) $\beta=9\,10^{-5}$, $f=.41$;\quad d) $\beta=9.75\,10^{-5}$, $f=.32$;\quad e) $\beta=9\,10^{-5}$, $f=.25$;\quad f) $\beta=13\,10^{-5}$, $f=.3$.}
  \label{fig:fre-res-beta}
\end{figure}

\begin{figure}[!t]
  \centering
  \includegraphics{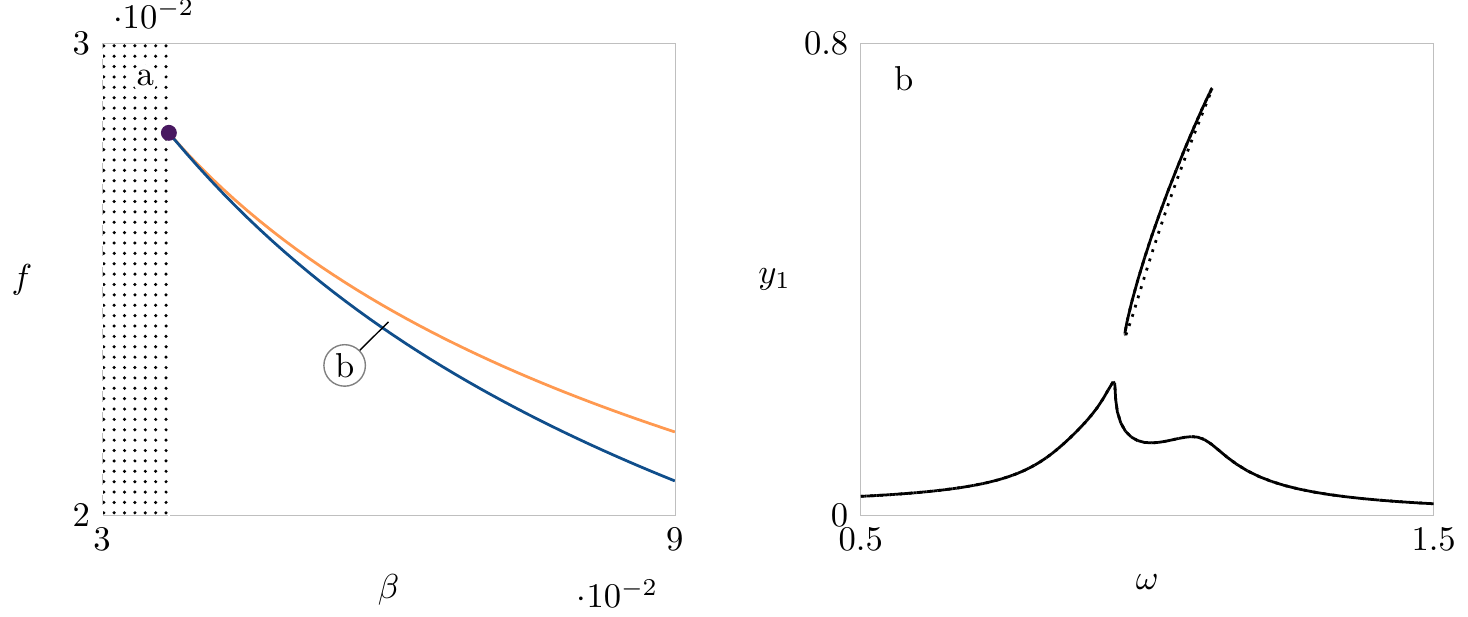}
  \caption{Asymmetric cusp for $\beta=3.7\,10^{-2}$ and corresponding DRC in the response.}
  \label{fig:fre-res-beta2}
\end{figure}

The influence of the fifth order parameter $\beta$ is illustrated in Fig.~\ref{fig:fre-res-beta}. The analysis reveals three distinct regions in which DRCs are present, each organised by an asymmetric cusp. Note that only one region exists in the absence of the fifth order spring ($\beta=0$). The significant outcome of the analysis for design purposes is that no DRC exists for $\beta>\beta_2\approx 1.2 \,10^{-4}$. A new asymmetric cusp appears at $\beta=\beta_3\approx 3.7 \,10^{-2}$ (fig.~\ref{fig:fre-res-beta2}), providing a range of parameter $[\beta_2,\beta_3]$ where the model is free of DRCs.

\begin{figure}[!h]
  \centering
  \includegraphics{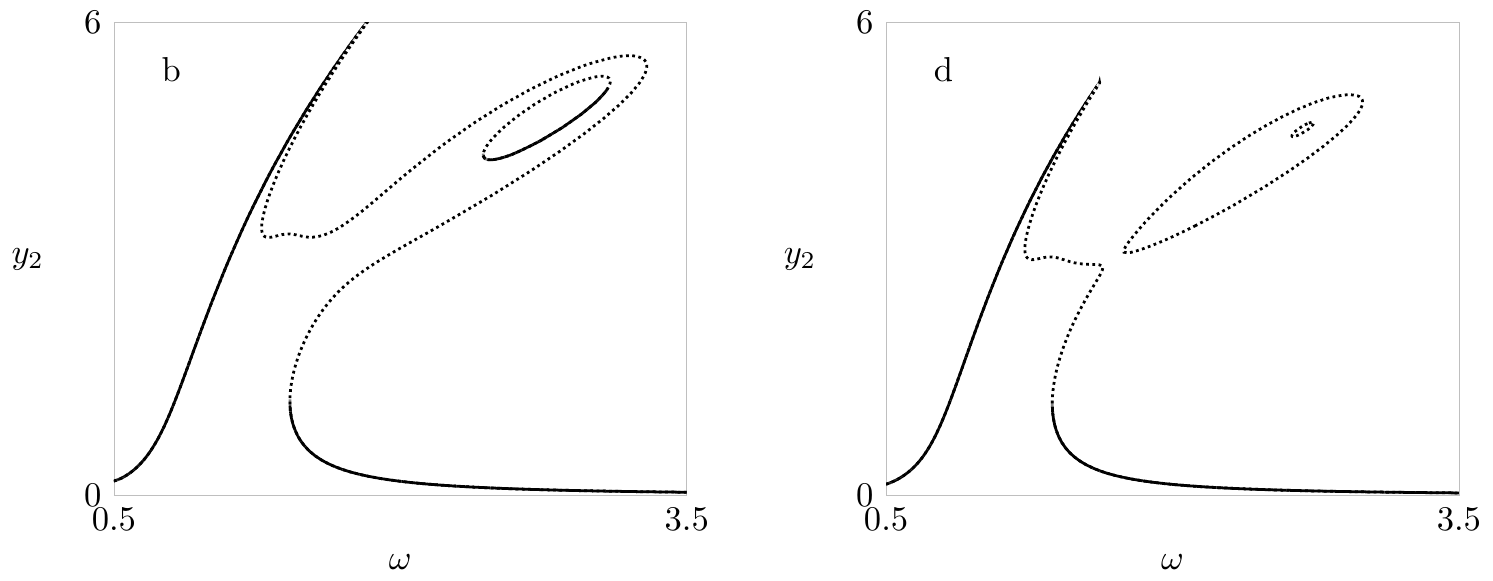}
  \caption{Presence of a new DRC due to a fifth order spring, amplitude responses of the second mass for the cases b and d in Fig.~\ref{fig:fre-res-beta}.}
  \label{fig:x2fr}
\end{figure}

Figure~\ref{fig:fre-res-beta} also shows the presence of another asymmetric cusp ($\beta_1\approx 10^{-4}$) and corresponding DRCs for $\beta<\beta_1$, for completeness Fig.~\ref{fig:x2fr} shows this other DRC in the response of the second mass.

All the frequency responses shown were computed using a shooting algorithm similar to that used in~\cite{Peeters2009nonlinear} (not from the equation obtained in~\ref{sec:HBNLTVA}). These figures confirm that the boundaries shown in Fig.~\ref{fig:fre-res-beta} are effectively present in the frequency response. Furthermore, numerical results show that the parameter values marking the boundaries differ only slightly from those obtained with first-order harmonic balance.

Stability properties are all included in the responses. In many cases the solutions are unstable, a fact that suggests the presence of quasiperiodic motions (cf.~\cite{Habib2015nonlinear,Detroux2015performance}), not considered in this work.

\section{Conclusions}
\label{sec:conclusions}

This paper illustrates the role of singularity theory in nonlinear resonance analysis and how it complements classical continuation methods. The theory was applied to the frequency response of a two-degree-of-freedom mechanical system. We showed that codimension one singularities characterise the transitions between qualitatively distinct regions of the frequency response. We also showed that the analysis of asymmetric cusps in this model provides a full characterisation of detached resonance curves.

Although the numerical approach in this paper was elementary, the results demonstrate the potential of this approach, and suggest that further developments in this direction could lead to valuable tools in the study of nonlinear resonance in mechanical systems.

\section{Acknowledgements}
\label{sec:acknowledgements}
G. Habib would like to acknowledge the financial support of the Belgian National Science Foundation FRS-FNRS (PDR T.0007.15). The research leading to these results has received funding from the European Research Council under the Advanced ERC Grant Agreement Switchlet n.670645

\appendix
\renewcommand*{\thesection}{\appendixname\,\Alph{section}}
\section{Reduction of first-order harmonic-balance equation for NLTVA}\label{sec:HBNLTVA}
In this section the harmonic-balance method with one harmonic is applied to system~\eqref{eq:NLTVAeom} with a fifth-order spring, and the resulting system is reduced to a single equation. The equations of motion are
\begin{equation}\label{eq:NLTVAeomApp}
    \begin{gathered}
      M \ddot{y}+C\dot{y}+Ky+f_{nl}(y)=g \frac{f}{2} (\exp(\i\omega t)+\exp(-\i\omega t)),\\[2ex]
      M=\left( \begin{array}{cc}m_1 & 0 \\ 0 & m_2\end{array}\right),\,\,\,\,\,\,K=\left(\begin{array}{cc}k_1+k_2 & -k_2\\ -k_2 & k_2\end{array}\right),\,\,\,\,\,\,C=\left(\begin{array}{cc}c_1+c_2 & -c_2 \\ -c_2 & c_2\end{array}\right),\\[2ex]
      f_{nl}(y)=\left(\begin{array}{c}\alpha_1y_1^3+\alpha_2(y_1-y_2)^3+\beta(y_1-y_2)^5\\\alpha_2(y_2-y_1)^3+\beta(y_2-y_1)^5\end{array}\right),\,\,\,\,\,\,g=\left(\begin{array}{c}1\\0\end{array}\right).
    \end{gathered}
\end{equation}
Defining
\begin{equation}
q=Ly,\,\,\,\,\,\,L=\left(\begin{array}{cc} 1 & 0\\1 & -1 \end{array}\right),
\end{equation}
the equations of motion are rewritten as
\begin{align}
  L^TML \ddot{q}+L^TCL\dot{q}+L^TKLq+L^Tf_{nl}(Lq)&=L^Tg \frac{f}{2} (\exp(\i\omega t)+\exp(-\i\omega t)),\\
  \phantom{=}&\phantom{=}\nonumber\\
  \begin{split}\label{eq:NLTVAeomq}
    (m_1+m_2)\ddot{q}_1-m_2 \ddot{q}_2+k_1q_1+c_1 \dot{q}_1+\alpha_1q_1^3&=\frac{f}{2} (\exp(\i\omega t)+\exp(-\i\omega t)),\\
    -m_2 \ddot{q}_1+m_2 \ddot{q}_2+k_2q_2+c_2 \dot{q}_2+\alpha_2  q_2^3+\beta q_2^5&=0.
  \end{split}
\end{align}
Setting $q_1=w_1\exp(\i\omega t)+\overline{w_1}\exp(-\i\omega t)$, $q_2=w_2\exp(\i\omega t)+\overline{w_2}\exp(-\i\omega t)$ and injecting in~\eqref{eq:NLTVAeomq}, yields for the first harmonic
\begin{alignat}{2}
  g_1(w_1,w_2)&:={}& \ctrmcell{ -\omega^2 (m_1+m_2) w_1 + \omega^2 m_2 w_2 + k_1 w_1 + \i c_1 \omega w_1 +  3\alpha_1 w_1^2 \overline{w_1}}  &= \frac{f}{2},  \label{eq:HBnltvaCom1} \\
  g_2(w_1,w_2)&:={}& \ctrmcell{ \omega^2 m_2 w_1 - \omega^2 m_2 w_2 + k_2 w_2 + \i c_2 \omega w_2 + 3 \alpha_2 w_2^2 \overline{w_2} + 10 \beta w_2^3\overline{w}_2^2} &= 0.   \label{eq:HBnltvaCom2}
\end{alignat}
Equation~\eqref{eq:HBnltvaCom2} can be solved for $w_1$, yielding
\begin{equation}
  \label{eq:w1val}
   w_1 = \psi(w_2) :=  w_2 - \frac{k_2}{\omega^2 m_2} w_2 - \i \frac{c_2}{\omega m_2} w_2 - \frac{3 \alpha_2}{\omega^2 m_2} w_2^2 \overline{w_2}  -\frac{10\beta}{\omega^2 m_2}w_2^3\overline{w}_2^2.
\end{equation}
$g_1(w_1,w_2)$ and $\psi(w_2)$ verify, for $\theta\in\mathbb{R}$
\begin{equation}
  \label{eq:equivariance}
  g_1(\exp(\i\theta)w_1,\exp(\i\theta)w_2)=\exp(\i\theta)g_1(w_1,w_2), \qquad \psi(\exp(\i\theta)w_2)=\exp(\i\theta)\psi(w_2).
\end{equation}
Thus for the equation obtained by substituting~\eqref{eq:w1val} in~\eqref{eq:HBnltvaCom1}
\begin{equation}
  \label{eq:quasiFinalEq}
  g_1(\psi(w_2),w_2)=\frac{f}{2},
\end{equation}
a similar property holds, i.e.
\begin{equation}
  \label{eq:equivariance2}
  g_1(\psi(\exp(\i\theta)w_2),\exp(\i\theta)w_2)= g_1(\exp(\i\theta)\psi(w_2),\exp(\i\theta)w_2)=\exp(\i\theta)g_1(\psi(w_2),w_2), \qquad \theta\in\mathbb{R}.
\end{equation}
Substituting $w_2=x \exp(\i\phi)$ in~\eqref{eq:quasiFinalEq},  multiplying by $\exp(-\i\phi)$ and using~\eqref{eq:equivariance2} yields
\begin{equation}
  \label{eq:NLTVAcvFe}
  g_1(\psi(x),x)=\frac{f}{2}\exp(-\i\phi),
\end{equation}
where $x$ and $f$ are real variables, but $g_1$ is complex-valued. Taking the square of the absolute value of both sides of~\eqref{eq:NLTVAcvFe} results in a real equation
\begin{equation}
  \label{eq:NLTVAaeApp}
  g:=\left|g_1(\psi(x),x))\right|^2-\tfrac{1}{4}f^2,
\end{equation}
whose solutions correspond to solutions of~\eqref{eq:HBnltvaCom1} and~\eqref{eq:HBnltvaCom2}. 

Furthermore, due to the structure of $g_i$
\begin{equation}
  \label{eq:gi_form}
  \begin{split}
    g_1 &= w_1h_1(|w_1|^2) + \omega^2 m_2 w_2\\
    g_2 &= w_2h_2(|w_2|^2) + \omega^2 m_2 w_1,
  \end{split}
\end{equation}
the resulting equation~\eqref{eq:NLTVAaeApp} depends quadratically on $x$ and $f$ in~\eqref{eq:NLTVAaeApp}, thus it is possible to substitute $X=x^2$ and $F=f^2$.

The frequency response of the system without fifth-order spring is obtained setting $\beta=0$.

\section*{References}
\label{sec:references}

\bibliographystyle{model1-num-names}
\bibliography{st_fr}

\end{document}